\documentclass[12pt]{article}
\usepackage{graphicx}
\begin{document}
\begin{center}
{\Large M-Theory on a Supermanifold}
\vskip 0.3 true in
{\large J. W. Moffat}
\date{}
\vskip 0.3 true in
{\it Department of Physics, University of
Toronto, Toronto, Ontario M5S 1A7, Canada}
\end{center}

\begin{abstract}%
A conjectured finite M-theory based on eleven-dimensional
supergravity formulated in a superspace with a non-anticommutative
$\diamondsuit$-product of field operators is proposed. Supermembranes are
incorporated in the superspace $\diamondsuit$-product formalism. When the deformed
supersymmetry invariant action of eleven-dimensional supergravity theory is expanded
about the standard supersymmetry invariant action, the spontaneously compactified
M-theory can yield a four-dimensional de Sitter spacetime inflationary solution with
dark energy described by the four-form F-fields. A fit to the present
cosmological data for an accelerating universe is obtained from matter
fields describing the dominant dark matter and the four-form F-field dark energy.
Chiral fermions are obtained from the M-theory by allowing singularities in the
compact internal seven-dimensional space. The possibility of obtaining a realistic
M-theory containing the standard model is discussed.
\end{abstract}  \vskip 0.2
true in {\it The great tragedy of Science -- the slaying of a beautiful hypothesis
by an ugly fact.}  \vskip 0.2 true in Thomas Huxley \vskip 0.2 true in

e-mail: john.moffat@utoronto.ca

\section{\bf Introduction}

For almost thirty years the idea that there exists a supersymmetry
between fermion and boson particles has been one of the basic motivations
for constructing a unified theory of gravitation and particle interactions.
This culminated in the formulation of string theories and their dual
relationships and M-theory, from which the five basic string theories
emanate by compactification~\cite{Townsend,Polchinski}.

The description of the fermions of the standard model with
the group structure $SU(3)\times SU(2)\times U(1)$ in
Kaluza-Klein and supergravity theories became problematic due to
Witten's application of the Atiyah-Hirzbruch chiral index
theorem~\cite{Atiyah,Witten,Wetterich,Zwiebach,Moffat}. The
possibility of using a smooth internal space with continuous
isometries was prevented by a no-go theorem, and spelt the end
of interest in a conventional interpretation of Kaluza-Klein
theories and eleven-dimensional supergravity theory. Superstring
theory in ten dimensions now came to the forefront and began to
dominate the particle physics and unification scene, since the
compactifications of string theory did not suffer the fate of
the chiral index no-go theorem, and promised to resolve the
problem of the consistency of quantum gravity by producing
finite calculations of graviton scattering amplitudes.

The advent of
duality theorems and the discovery that an eleven-dimensional
M-theory with N=1 supersymmetry should constitute the fundamental
unified theory with string theory evolving from a
compactification of the eleventh dimension, raised again the
question as to why nature would stop at ten dimensions in its
use of supersymmetry. The maximum supersymmetry was obtained in
eleven dimensions -- not ten -- as in string theory. Beyond
eleven dimensions, there ceased to be a supersymmetry and if one
entertained two time dimensions severe causality problems
would ensue. M-theory has eleven-dimensional
supergravity theory as a low energy limit with its associated
type IIA string theory and mass spectrum, so eleven-dimensional
supergravity rose like the `phoenix from the ashes' and became
again a subject of intensive research.

Now the questions arise: What constitutes an M-theory? How many
degrees of freedom does M-theory possess? One important fact
must be considered, namely, that {\it there are no strings} in
eleven-dimensional M-theory, because the only boson components
of the action are the three-form potential $A_{MNP}$ where
$M,N,P=0,1...10$ and the four-form $F_{MNPQ}$ obtained from the
curl of $A_{MNP}$. So we cannot rely on the finiteness of string
theory to guarantee a finite eleven-dimensional M-theory.
However, supermembranes can couple to $A_{MNP}$ and $F_{MNPQ}$,
so these higher-dimensional objects can play a fundamental role
in eleven dimensions. But such branes cannot be quantized in
perturbation theory, because the standard method of string
perturbation theory quantization does not work. Each term in
string perturbation theory corresponds to a two-dimensional
worldsheet with an increasing number of holes. A sum over all
topologies of the world sheet is performed. But for surfaces
with more than two dimensions-- as encountered with branes-- we
do not know how to do this. Powerful theorems in mathematics
preclude the possibility of doing this, since we are unable to
classify surfaces with $p > 3$. Attempts have been made
to quantize the supermembrane in a non-covariant light
front gauge~\cite{Duff}. The results indicate that the
supermembrane has a continuous spectrum leading to instability.

Treating branes
as fundamental objects raises the unpleasant question of how to
perform non-perturbative quantization. So the issue of how to
implement {\it finiteness} of calculations of amplitudes in
M-theory and quantum gravity has returned in a new guise. The
question arises whether the resolution of the finiteness problem
in eleven dimensions would lead to new physics beyond string
theory on compactification of the correct M-theory.

Ho\u{r}ava and Witten~\cite{Horava} performed an orbifold
compactification of M-theory along a finite length of the
eleventh dimension and obtained a ten-dimensional bulk with two
(9+1) branes, each with an $E8$ heterotic string gauge
structure. It was shown that this picture could be compacitfied
to a five-dimensional bulk-brane scheme and this led to the
phenomenological Randall-Sundrum scenario~\cite{Sundrum}. Due to
the existence of boundaries in the bulk world and associated
discontinuities at the junction of bulk-brane, the chiral no-go
theorem was avoided. However, we are still left with the
unanswered question of the meaning of the original
eleven-dimensional M-theory.

One suggestion for an
M-theory was to introduce noncommuting eleven-dimensional
coordinates in the form of $N\times N$ matrices and construct a
Lagrangian with noncommuting coordinates in a light-front gauge
and then take the $N\rightarrow\infty$ limit to obtain a super
Yang-Mills gauge structure~\cite{Susskind}. However, this theory is tied to the
specific light-front gauge and it is not clear how it produces a
finite eleven-dimensional theory. A rigorous proof that it
produces a gauge-free eleven-dimensional supergravity theory at
low energies is lacking.

We shall propose that M-theory is structured on an
eleven-dimensional supermanifold of superspace coordinates with
both commuting `boson' and anticommuting `fermion'
Grassmann
coordinates~\cite{Moffat2,Moffat3,Moffat4,Moffat5}.
An associated operator Hilbert space has noncommuting and
non-anticommuting coordinates and we map this operator space to
the space of ordinary superspace coordinates and fields by a
$\circ$-product of field operators or a $\diamondsuit$-product
of field operators. The eleven-dimensional supergravity action
is now invariant under a generalized `deformed' group of
supersymmetry transformations. This formalism leads to a finite
and unitary perturbation theory in eleven dimensions. The
noncommutative quantum field theory, based on the familiar
`bosonic' commuting coordinates $x^M$, cannot by itself lead to
a renormalizable or finite M-theory.

Recent observations of supernovae and the cosmic microwave background have led to the
mounting evidence that the universe is presently undergoing an accelerating
expansion~\cite{Perlmutter,Netterfield}. This means that either there is a positive,
small cosmological constant or a slowly varying quintessence field energy (dark
energy)~\cite{Caldwell} that dominates the present universe. This has led to a
crisis in string theory and M-theory, since it is difficult to invoke a positive
cosmological constant and the associated de Sitter (dS) spacetime in supersymmetric
theories that produce stable and consistent solutions. This is particularly true for
eleven-dimensional M-theory and supergravity and ten-dimensional string theory.
Moreover, eternal quintessence acceleration and dS spacetimes lead to future
cosmological event horizons which prevent the construction of an S-matrix theory --
a serious problem for string theory. In early constructions of eleven-dimensional
supergravity, it was realized that supersymmetric vacua preferred a spontaneous
compactification resulting in a four-dimensional anti-de Sitter (AdS) spacetime and
a compact internal seven-dimensional compact
space~\cite{Freund,Freund2,Englert,Duff2}. This compacitification naturally selected
the dimensions of spacetime to be four.

A spontaneous compactification of our M-theory to a product space
$M_4\times M_7$, where $M_4$ is a maximally symmetric
Friedmann-Robertson-Walker spacetime and $M_7$ is a compact
Einstein space, can produce solutions of the cosmological field
equations which possess AdS and dS
solutions in four dimensions as well as a solution corresponding
to a zero four-dimensional and seven-dimensional cosmological
constant. We postulate that the four-form field
strengths of supergravity describe the `dark energy' associated with quintessence
and the vacuum energy.

When we take the limit of standard
eleven-dimensional supergravity and its associated supersymmetry
transformations, we obtain on compactification the familiar
result of the product of an AdS spacetime and a positively
curved seven-dimensional Einstein space. By including matter fields obtained from a
compactification of our M-theory to four dimensions and the vacuum dark energy
solution with a small positive cosmological constant $\lambda_4$, we can obtain a fit
to the present cosmological data describing an accelerating universe.

An old problem in eleven-dimensional supergravity is
obtaining a satisfactory isometry group on compactification
that can include the standard model $SU(3)\times SU(2)\times
U(1)$. In the next section, we shall review the problem of obtaining chiral fermion
representations and non-Abelian gauge fields from compactifications of
eleven-dimensional M-theory.

\section{\bf The Chiral Fermion Problem}

Let us review the origin of the chiral fermion no-go theorem. Compact
internal spaces are characterized by an index~\cite{Atiyah} which
determines the number of chiral fermions. The nature and application of
this index in the context of dimensional reduction has been discussed by
Witten and Wetterich~\cite{Witten,Wetterich} and in the context of
non-Riemannian geometry in ref.~\cite{Moffat}. Consider an
operator $D$ acting on spinors containing first derivatives of
the spinor and non-derivative terms and that this operator
commutes with existing gauge transformations and anticommutes
with $\Gamma_d$, the d-dimensional internal space gamma matrix.
The chiral index is defined by
\begin{equation}
N_C(D)=n^+_C-n^-_C-n^+_{\bar C}+n^-_{\bar C},
\end{equation}
where $n^+_C$ denotes the number of zero modes in the
$d$-dimensional internal space of the Weyl spinor $\psi^+$
belonging to the complex representation $C$ and with the same
kind of meaning for $n^-_C,n^+_{\bar C}$ etc. We suppose that we
perform a dimensional reduction on the $(d+4)$ space to a Dirac
operator ${\cal D}$, so that $D$ is the mass operator and $N_C$
counts the number of chiral four-dimensional left-handed fermion
generations in the $C$ representation.

We denote by $e^A_M$ the D-dimensional vielbein where $M,N$ denote the
world space coordinates and $A,B$ the tangent space coordinates. If the
viebein has an inverse
\begin{equation}
e^M_Ae^B_M=\delta^B_A,\quad e^A_Me^N_A=\delta^N_M,
\end{equation}
then we have
\begin{equation}
\Gamma^MD_M\psi=g^{MN}e_{NA}\Gamma^AD_M\psi,
\end{equation}
where the metric $g_{MN}$ may also have an inverse
\begin{equation}
g_{MN}g^{NP}=\delta_M^P.
\end{equation}
The internal d-dimensional Dirac operator is
\begin{equation}
\Gamma^mD_m\psi=g^{\Sigma\Delta}e_{\Delta m}\Gamma^mD_\Sigma\psi.
\end{equation}

If the vielbein $e^m_\Delta$, where $m,n$ and $\Delta,\Sigma$
denote the internal tangent space and space indices,
respectively, is invertible everywhere, then the operator
$\Gamma^\Delta D_\Delta$ is an elliptic operator. The index $N_C$
remains invariant under continuous changes of the metric, the
vielbein and the connections for elliptic operators that
preserve the isometries of the space. On compact spaces we
have~\cite{Witten}
\begin{equation}
N_C(D)=N_C(D+aO).
\end{equation}
Here, $a$ is arbitrary and $O$ is an operator satisfying
\begin{equation}
[{\cal G},O]=0,\quad \{\Gamma_d,O\}=0,
\end{equation}
where ${\cal G}$ denotes the generators of the isometry group. For
everywhere invertible vielbeins the chiral index $N_C(D)=0$ and
spinors have a vector-like behaviour in the compact internal
space.

We must allow for singularities in our internal compact space in order to
obtain chiral fermions on dimensional reduction. This has led recently to a
study of $G_2$ holonomy spaces which contain conical
singularites~\cite{Acharya,Witten2,Witten3}.

There are no complex spinor representations in an odd-dimensional space, and in
particular in the eleven-dimensional space of M-theory. The chiral fermions must
arise from a compactification of the eleven-dimensional M-theory
space $M_{11}=M_4\times M_7$. For the choice $M_7=S^7$, we
cannot permit a singularity within the smooth seven sphere $S^7$.

If we gauge the
supergravity internal space by allowing for non-compact groups such as
$SO(p,r)$, then we find that there can exist singular points in the group space
manifold and the vielbeins are not invertible at these singular points, leading to a
non-vanishing chiral index and chiral fermions. These possibilities were recognized
soon after the discovery of the no-go
theorems~\cite{Wetterich,Zwiebach,Moffat} but they were not pursued in
earnest, because superstrings again became the subject of dominant interest. The
non-compact versions of internal spaces suffer from instabilities and the lack of a
physical mass spectrum.

Superstrings have extra gauge fields within their
higher-dimensional spaces and these automatically guarantee a non-vanishing chiral
index $N_C$. Indeed, conventional Kaluza-Klein theories can be extended to include
extra gauge fields~\cite{Salam} but such models were considered unattractive,
because they ruined the pure higher-dimensional gravity approach of Kaluza-Klein
theories and supergravity theories.

The proposed solution of the chiral fermion problem in M-theory by Acharya and
Witten~\cite{Acharya} compactifies M-theory on a seven manifold $X$ of $G_2$
holonomy that preserves the supersymmetric ground state. It leads naturally to
a four-dimensional theory with $N=1$ supersymmetry. When the manifold $X$ is smooth,
then one obtains Abelian gauge groups only without chiral fermions. However, the
singular manifolds of $G_2$ holonomy offer the possibility of generating non-Abelian
gauge groups and chiral fermions. In codimension six, M-theory singularities can
occur in Calabi-Yau threefolds, which can be embedded in a manifold of $G_2$
holonomy. There are no known complete classifications of $G_2$ holonomy spaces.
Nevertheless, the known isolated singularities of $G_2$-manifolds are conical
singularities, and some of them generate chiral fermions and certain special models
of this kind lead to anomaly cancellation. Acharya and Witten~\cite{Acharya} have
investigated examples constructed by taking the quotient of a conical hyper-Kahler
eight-manifold by a $U(1)$ symmetry group. The compactification to four-dimensional
spacetime leads to a flat Minkowski spacetime with vanishing cosmological constant.
At present, it is not known how to compactify a seven manifold to a four-dimensional
de Sitter spacetime with chiral fermions and non-Abelian gauge fields.

\section{\bf The Accelerating Universe and de Sitter Spaces}

Much of the research in particle physics over the past thirty years has
focussed on fundamental interactions and their unification in relation to
supersymmetry. This is certainly true for string theory and M-theory. We
know that supersymmetry must be badly broken for energies less than about
$1-10$ TeV, so we are already confronted with the need to break
supersymmetry at low energies. There is no well-defined theory for doing
this and this has been a thorn in the side of string theory research for a
long time. The issue of supersymmetry breaking has been exacerbated
recently with the cosmological observations based on supernovae and of the
anisotropy of the cosmic microwave background, which suggest that the
universe entered a stage of accelerated expansion a few billion years after
the big bang~\cite{Perlmutter,Netterfield}. Standard inflationary models are based on
the idea that the energy-momentum tensor is dominated by the potential energy of a
scalar field, $V(\phi)$, with $V > 0$~\cite{Linde}. The scalar field is assumed to
be slowly rolling with a flat potential with $\dot{\phi}^2 \ll V(\phi)$ and the
limiting case ${\dot\phi}=0$ corresponds to a cosmological constant with $V > 0$.
When M/string theory is compactified on an internal compact space, then a
supersymmetric vacuum will produce a stable AdS (3+1) space (with negative
cosmological constant) and an Einstein seven-dimensional space with positive
curvature. When we attempt to extend these solutions to (3+1) dS spacetimes and
Einstein spaces with negative curvature, then the solutions are unstable for
supersymmetric vacua and even for broken supersymmetric
vacua~\cite{Gibbons,Maldacena}.

Recently, investigations have centered on obtaining stable dS spacetimes in
extended four-dimensional supergravity theories. Whereas such solutions may
be obtainable, they do not help us in making progress in string theory or
higher-dimensional unification theories such as M-theory. This has led to a
renewed interest in studying gauged higher-dimensional supergravity
theories with non-compact internal spaces and finite
volume~\cite{Hull}. However, such theories immediately run into
serious difficulties. To be physically realistic and correspond
to a stable universe, they must possess a discrete mass spectrum
with a mass gap. Such internal non-compact spaces have been
studied by Nicolai and Wetterich~\cite{Nicolai}, who found that
it is possible to obtain normalizable wave functions associated
with a discrete spectrum when appropriate boundary conditions
are imposed. However, a more serious problem arises due to the
kinetic energies of scalar and antisymmetric gauge fields
possessing the {\it wrong sign}, as is to be expected for
non-compact spaces. These anomalous kinetic terms violate
unitarity and cause the theories to be unstable and unphysical.
One might argue that at the scale of the Hubble distance
$H_0^{-1}$, such violations of unitarity can be ignored but this
is courting trouble! What is inherently unphysical will stay to
haunt the innocent researcher and invariably lead to unpleasant
unphysical behaviour.

\section{\bf Complementary Chiral Symmetry and de Sitter Space Problems}

In our search for a consistent unified description of fundamental forces
including gravity based on an M-theory in eleven dimensions, we are
currently caught in a dilemma. To guarantee a physical chiral
fermion mass spectrum, we would prefer to compactify the M-theory
spontaneously within a compact space with orbifold singularities or
isolated singularities such as conical singularities. However, when we
attempt to obtain stable dS solutions within a supersymmetric
scheme or within a broken supersymmetric framework (dS solutions
in M-theory would break most or all of the supersymmetry), then
we fail to do so within a conventional compact internal space
compactification. We could avoid the chiral index no-go theorem
by attempting gauged supergravity models with non-compact
internal group structures, thereby, finding dS and
inflationary solutions but these may still not be stable for
broken supersymmetry ground states and they suffer from serious
unitarity violations with their related instabilities.

There is still the third problem that if we succeed in deriving a
stable dS solution in M-theory with chiral fermions associated
with the compactification, then we are faced with potential
generic future cosmological horizons which make any attempt to
formulate a physical S-matrix doomed to failure~\cite{Bousso,Fischler}.
Since string theory is a first quantized S-matrix theory and any
conceivable M-theory formulated within scenarios based on modern
physics is only sensibly formulated within standard S-matrix
theory, then we are confronted with an uncontrollable foundation
for the theory of everything!

The prospect of having to formulate a theory of quantum gravity in de
Sitter space is unpleasant to say the least. It is hard to define
meaningful physical operators in de Sitter space, there is no unique vacuum
in de Sitter space and there is the problem of constructing a field theory
with only a finite number of degrees of freedom. A glance at a possible
quantum gravity theory based on perturbation theory, shows that spin 2
propagators are unwieldy in x-space and it is hard to make sense of Green's
functions in
momentum space, for Fourier transforms have to be performed on
operator spaces that do not possess a time-like or null-like
infinity and the operators do not form a complete set of
observables in the usual sense. Much of the progress in
physics since Newton, has been made using some form of
perturbation theory. Little or no progress has been made in
studies of non-perturbative quantum field theory. However, it
might seem that ultimately a non-perturbative route towards a workable
M-theory is needed.

\section{Supermanifolds}

In general, we can parameterize the superspace coordinates as
$\rho^{\bar M}=(x^y,\beta^\alpha)$ where the $x^y$ are the commuting
spacetime coordinates that belong to a smooth topological manifold and
the $\beta^\alpha$ is an anticommuting spinor field,
$\{\beta^\alpha,\beta^\sigma\}=0$.

Associated with our supervector space ${\cal S}$ is the ${\cal
S}$-parity mapping ${\cal P}:{\cal S}\rightarrow {\cal S}$ acting
on an arbitrary supervector ${\vec X}={\vec X}_c+{\vec X}_a$ by
the rule~\cite{Buchbinder,Berezin}:
\begin{equation}
{\cal P}({\vec X})={\vec X}_c-{\vec X}_a.
\end{equation}
For a pure supervector ${\vec X}$, we have
\begin{equation}
{\cal P}({\vec X})=(-1)^{\epsilon({\vec X})}{\vec X},
\end{equation}
where $\epsilon({\vec X})=0$ for a c-number commuting supervector
and $\epsilon({\vec X})=1$ for an anticommuting supervector.

For real superspaces $\Re^{p,q}$ where $p$ and $q$
denote the dimensions of the $x$ and $\beta$ coordinates, we can
introduce general coordinate transformations. Consider the
one-to-one mapping of $\Re^{p,q}$ on itself. This is
\begin{equation}
\rho^M\rightarrow\rho^{'M}=f^M(\rho).
\end{equation}
We shall restrict ourselves to supersmooth transformations and
the condition for an invertible transformation is that the
supermatrix
\begin{equation}
\label{transf}
H_M^N=\partial_Mf^N(\rho)
\end{equation}
is non-singular at each point $\rho^M\in \Re^{p,q}$. The set of
all invertible supersmooth transformations (\ref{transf}) forms
a group called the supergroup of diffeomorphism transformations.
We can also have superalgebras ${\cal A}$ which are
$Z_2$-graded linear spaces ${\cal S}={\cal S}_c\oplus {\cal
S}_a$ with the multiplication law
\begin{equation}
[A,B\}=A\cdot B-(-1)^{\epsilon(A)\epsilon(B)}B\cdot A,
\end{equation}
where $A$ and $B$ are arbitrary pure elements and $[...,...\}$
is the graded Lie superbracket. The super-Jacobi identities are
\begin{equation}
(-1)^{\epsilon(A)\epsilon(C)}[A,[B,C\}\}+(-1)^{\epsilon
(B)\epsilon(A)}[B,[C,A\}\}+(-1)^{\epsilon(C)\epsilon(B)}[C,[A,B\}\}=0.
\end{equation}
For arbitrary elements $A,B$ and $C$ we have
\begin{equation}
[A,B\}=[A_c,B_c]+[A_c,B_a]+[A_a,B_c]+\{A_a,B_a\}.
\end{equation}

For super Lie algebras ${\cal A}$ one can connect a super Lie
group ${\cal G}$ by the exponential rule: ${\bf
g}(\xi^i)=\exp({\bf a})$ where ${\bf a}=\xi^ie_i$, $e_i$ is
the basis for ${\cal G}$, ${\bf a}$ are the super Lie algebra
elements and the $\xi^i$ ($i=1,2,..., {\rm dim}\, {\cal G}$) are
the components of the Lie algebra elements which play the role of
local coordinates in the neighborhood of the identity of the
group manifold~\cite{Berezin}. We have a generalized Poincar\'e
supergroup and superalgebra which are extensions of the
corresponding standard Poincar\'e group and algebra.

\section{\bf M-Theory Based on Noncommutative and Non-Anticommutative Geometry}

If we believe that there exists a physical, unified description of fundamental
interactions described by an eleven-dimensional M-theory, then we must seek
an eleven-dimensional structure that produces a finite M-theory,
compactifies to a (3+1) spacetime and a seven-dimensional space with chiral
fermions and contains solutions to the field equations which describe a
stable de Sitter space. Such a scenario should lead to a finite quantum
gravity in four dimensions and give a realistic description of cosmology
and contain the standard model of quarks and leptons. The superspace
contemplated can allow a generalization of the standard supersymmetry
transformations associated with eleven-dimensional supergravity and a
generalized supersymmetric ground state. These new supersymmetry
transformation laws can be considered as deformations of the standard
classical supersymmetry transformations. We shall show that such a
generalization of the supersymmetric invariance group of supergravity can
allow four-dimensional de Sitter solutions.

Instead of pursuing a
non-perturbative quantization of supermembranes in eleven dimensions, we
shall introduce a superspace with
supercoordinates~\cite{Moffat2,Moffat3,Moffat4,Moffat5}
\begin{equation}
\label{coordinates}
\rho^M=x^M+\beta^M,
\end{equation}
where the $x^M$ are eleven-dimensional commuting coordinates
and $\beta^M$ are Grassmann anticommuting coordinates, so they
are even and odd elements of a Grassmann algebra, respectively.

Both noncommutative and
non-anticommutative geometries can be unified within the
superspace formalism using the $\circ$-product of two operators
${\hat f}$ and ${\hat g}$~\cite{Moffat2,Moffat3,Moffat4,Moffat5}:
\begin{equation}
\label{fgproduct}
({\hat f}\circ{\hat g})(\rho)
=\biggl[\exp\biggl(\frac{1}{2}\omega^{MN}
\frac{\partial}{\partial\rho^M}\frac{\partial}{\partial\eta^N}\biggr)
f(\rho)g(\eta)\biggr]_{\rho=\eta} $$ $$
=f(\rho)g(\rho)+\frac{1}{2}\omega^{MN}\partial_M
f(\rho)\partial_Ng(\rho)+O(\vert\omega\vert^2),
\end{equation}
where $\partial_M=\partial/\partial\rho^M$ and $\omega^{MN}$ is a
nonsymmetric tensor
\begin{equation}
\omega^{MN}=\tau^{MN}+i\theta^{MN},
\end{equation} with
$\tau^{MN}=\tau^{NM}$ and $\theta^{MN}=-\theta^{NM}$.
Moreover, $\omega^{MN}$ is Hermitian symmetric
$\omega^{MN}=\omega^{\dagger MN}$, where $\dagger$ denotes Hermitian
conjugation. The familiar commutative coordinates of spacetime are replaced
by the superspace operator relations
\begin{equation}
\label{commutator}
[{\hat\rho}^M,{\hat\rho}^N]=2\beta^M\beta^N+i\theta^{MN},
\end{equation}
\begin{equation}
\label{anticommutator}
\{{\hat\rho}^M,{\hat\rho}^N\}=2x^Mx^N+2(x^M\beta^N+x^N\beta^M)+\tau^{MN}.
\end{equation}
In the limits that $\beta^M\rightarrow 0$ and
$\vert\tau^{MN}\vert\rightarrow 0$, we get the familiar expression for
noncommutative coordinate operators
\begin{equation}
[{\hat x}^M,{\hat x}^N]=i\theta^{MN}.
\end{equation}
In the limits $x^M\rightarrow 0$ and $\vert\theta^{MN}\vert\rightarrow 0$,
we obtain the Clifford algebra anticommutation relation
\begin{equation}
\{{\hat\beta}^M,{\hat\beta}^N\}=\tau^{MN}.
\end{equation}

We shall use the simpler non-anticommutative
geometry obtained when $\theta^{MN}=0$ to construct the M-theory,
because it alone can lead to a finite and unitary quantum field theory and
quantum gravity theory ~\cite{Moffat2,Moffat3,Moffat4,Moffat5}.
In the non-anticommutative field theory formalism, the product of two
operators ${\hat f}$ and ${\hat g}$ has a corresponding
$\diamondsuit$-product
\begin{equation}
\label{diamondproduct}
({\hat f}\diamondsuit {\hat g})(\rho)
=\biggl[\exp\biggr(\frac{1}{2}\tau^{MN}\frac{\partial}{\partial\rho^M}\frac{\partial}
{\partial\eta^N}\biggr)f(\rho)g(\eta)\biggr]_{\rho=\eta} $$ $$
=f(\rho)g(\rho)+\frac{1}{2}\tau^{MN}\partial_M f(\rho)\partial_N
g(\rho)+O(\tau^2).
\end{equation}
We choose as the basic action of the
M-theory, the CJS~\cite{Julia} action for
eleven-dimensional supergravity, replacing all products of field operators
with the $\diamondsuit$-product. The action is invariant under
the generalized $\diamondsuit$-product supersymmetry gauge
transformations of the vielbein $e^A_M$, the spin 3/2 field
$\psi_M$ and the three-form field $A_{MNQ}$. Our M-theory has as
a low energy limit the CJS supergravity theory when
$\vert\tau^{MN}\vert\rightarrow 0$ and $\beta^M\rightarrow 0$.

Although we derive many of our basic results for our field theory
formalism in the superspace configuration space, in practice all
calculations of amplitudes using our generalized Feynman rules
will be performed in momentum space. To this end, we must make a
simplifying ansatz that defines the Fourier transform of an
operator ${\hat f}(\rho)$:
\begin{equation}
{\hat f}(\rho)=\frac{1}{(2\pi)^D}\int d^Dp
\exp(ip\rho){\tilde f}(p).
\end{equation}
When this Fourier transform of the operator ${\hat f}(\rho)$ is
invertible, then we can calculate matrix elements in momentum
space without having to concern ourselves with additional volume
factors that can occur due to integrations over the Grassmann
coordinates $\beta$. These additional volume factors occur in
standard superspace integrations~\cite{Buchbinder}.

We have for the $\diamondsuit$-product the exponential
function rule
\begin{equation}
\exp(ik\rho)\diamondsuit\exp(iq\rho)=\exp[i(k+q)]\exp\biggl[\frac{1}{2}(k\tau
q)\biggr],
\end{equation}
where $k\tau q=k_M\tau^{MN}q_N$.
Then for the $\diamondsuit$-product
of two operators ${\hat f}$ and ${\hat g}$ we get
\begin{equation}
({\hat f}\diamondsuit{\hat g})(\rho)=\frac{1}{(2\pi)^{2D}}\int
d^Dkd^Dq{\tilde f}(k){\tilde g}(q)\exp\biggl[\frac{1}{2}(k\tau
q)\biggr]\exp[i(k+q)\rho],
\end{equation}
where
\begin{equation}
{\tilde f}(k)=\frac{1}{(2\pi)^D}\int
d^D\rho\exp(-ik\rho){\hat f}(\rho).
\end{equation}

It was demonstrated in previous work
~\cite{Moffat3,Moffat4,Moffat5}, that scalar quantum
field theory and weak field quantum gravity are finite to all orders of
perturbation theory, and the S-matrix for these theories is expected to
obey unitarity. The regularization of the ultraviolet divergence of the
standard local point-like quantum field theory is caused by an exponential
damping of the Feynman propagator. The modified Feynman propagator ${\bar\Delta}_F$
is defined by the vacuum expectation value of the time-ordered
$\diamondsuit$-product
\begin{equation}
\label{propagator}
i{\bar\Delta}_F(\rho-\eta)\equiv\langle 0\vert
T(\phi(\rho)\diamondsuit(\eta))\vert 0 \rangle
$$ $$
=\frac{i}{(2\pi)^D}\int\frac{d^Dk\exp[-ik(\rho-\eta)]\exp[-\frac{1}{2}(k\tau
k)]}{k^2+m^2-i\epsilon}.
\end{equation}
Since we shall be performing all the
calculations of Feynman graphs in momentum space, the
limit $\rho\rightarrow\eta$ in (\ref{propagator}) is
not rigorously applied.

In momentum space we choose
the canonical value for the symmetric tensor
$\tau^{\mu\nu}=\delta^{\mu\nu}/\Lambda^2$ where $\Lambda$
denotes an energy scale. We shall set $\Lambda$ equal to the
eleven-dimensional Planck mass $M^{(11)}_{\rm PL}$ and since the
eleven-dimensional gravitational constant, $G^{(11)}$, is
proportional to Newton's constant i.e. the inverse of the Planck
mass squared, then our M-theory is free of arbitrary parameters.
We obtain in momentum space
\begin{equation}
i{\bar\Delta}_F(p)=\frac{i\exp\biggl(-\frac{p^2}{2\Lambda^2}\biggr)}
{p^2+m^2-i\epsilon},
\end{equation}
where $p$ denotes the Euclidean D-momentum vector. This
reduces to the standard commutative field theory form for the
Feynman propagator
\begin{equation}
i\Delta_F(p)=\frac{i}{p^2+m^2-i\epsilon}
\end{equation} in
the limit $\vert\tau^{\mu\nu}\vert\rightarrow 0$ and
$\Lambda\rightarrow\infty$.

An analysis of scattering amplitudes shows that due to
the numerator of the modified Feynman propagator being an entire function
of $p^2$, there are no unphysical singularities in the finite complex $p^2$
plane and the Cutkosky rules for the absorptive part of the scattering
amplitude are satisfied, leading to unitarity of the
S-matrix~\cite{Moffat4}.

An analysis of the higher-dimensional field theories,
including supersymmetric gauge theories generalized to the
non-anticommutative formalism, shows that they will also be finite to all
orders of perturbation theory. This result is mainly due to the
fundamental length scale $\ell$ in the theory that owes its existence
to a quantization of spacetime. When $\ell\rightarrow 0$
($\Lambda\rightarrow\infty$) the non-anticommutative field theories reduce
to the standard local point particle field theories which suffer the usual
difficulties of infinities and non-renormalizable quantum gravity.

Our finite field theory formalism is based on a nonlocal field theory, due to the
infinite number of derivatives in the exponential function that occurs in our
$\circ$-product or $\diamondsuit$-product of field operators. The nonlocal
modification of standard local point field theory occurs at short distances or high
energies where accelerator physics has not been probed.

It should be emphasized that the promise of significant progress in
noncommutative quantum field theory has not been
fulfilled~\cite{Szabo,Moffat6,Moffat7}. The planar Feynman graphs of perturbative
noncommutative scalar and Yang-Mills field theories, in which the actions are
described by Groenwald-Moyal $\star$-products of field operators, are as divergent
as in local point field theory, so that the ultraviolet behaviour of scattering
amplitudes is no better than in the standard renormalizable theories, albeit that
the non-planar graphs do exhibit an oscillatory damping of ultraviolet
divergences. This is also true of noncommutative weak field quantum
gravity~\cite{Moffat6,Moffat7} which continues to be nonrenormalizable. On the other
hand, the non-anticommutative quantum field theories in superspace are perturbatively
finite and unitary. However, all these quantum field theories are nonlocal at short
distances and further study of such nonlocal field theories is required.

We can now contemplate an M-theory which is finite to all orders of
perturbation theory and contains an invariance of the eleven-dimensional
supergravity under a {\it generalized supersymmetry transformation} with a
ground state that reduces to the standard supersymmetric ground state of
CJS in the limit that the energy scale $\Lambda\rightarrow\infty$.

\section{\bf Superspace M-Theory Action}

The field content consists of the vielbein $e^A_M$, a
Majorana spin $\frac{3}{2}$ $\psi_M$, and of a completely antisymmetric
gauge tensor field $A_{MNP}$. The metric is $(-+++...+)$ and the
eleven-dimensional Dirac matrices satisfy
\begin{equation}
\{\Gamma_A,\Gamma_B\}=-2\eta_{AB},
\end{equation}
where $\eta_{AB}$ denotes the flat Minkowski tangent space metric.
Moreover, $\Gamma^{A_1...A_N}$ denotes the product of $N\Gamma$ matrices
completely antisymmetrized.

Our superspace M-theory Lagrangian,
using the $\diamondsuit$-product has the
form~\cite{Julia,Nieuwenhuizen,Duff2}:
\begin{equation}
\label{Lagrangian} {\cal L_{\rm SG}}
=-\frac{1}{4\kappa^2}e\diamondsuit R(\omega)_{\diamondsuit}
-\frac{i}{2}e\diamondsuit{\bar\psi}_M\diamondsuit\Gamma^{MNP}D_N\biggl(\frac{\omega+{\hat\omega}}{2}\biggr)
\diamondsuit\psi_P-\frac{1}{48}e\diamondsuit F_{MNPQ}\diamondsuit F^{MNPQ}
$$ $$
+\frac{\kappa
}{192}e\diamondsuit({\bar\psi}_M\diamondsuit\Gamma^{MNOPQR}\psi_N+12{\bar\psi}^P\diamondsuit
\Gamma^{OR}\psi^Q)\diamondsuit(F_{PQOR}+{\hat F}_{PQOR})
$$ $$
+\frac{2\kappa}{(144)^2}\epsilon^{O_1O_2O_3O_4P_1P_2P_3P_4MNR}F_{O_1O_2O_3O_4}
\diamondsuit F_{P_1P_2P_3P_4}\diamondsuit A_{MNR}, \end{equation} where
$R(\omega)_{\diamondsuit}$ is the scalar contraction of the
curvature tensor
\begin{equation}
R_{MNAB}=\partial_M\omega_{NAB}-\partial_N\omega_{MAB}+{\omega_{MA}}^C\diamondsuit\omega_{NCB}
-{\omega_{NA}}^C\diamondsuit\omega_{MCB},
\end{equation}
and $F_{MNOP}$
is the field strength defined by
\begin{equation}
F_{MNOP}=4\partial_{[M}A_{NOP]},
\end{equation}
with $[...]$ denoting the
antisymmetrized sum over all permutations, divided by their number.

The covariant derivative is
\begin{equation}
D_N(\omega)\psi_M=\partial_N\psi_M+\frac{1}{4}\omega_{NAB}
\diamondsuit\Gamma^{AB}\psi_M.
\end{equation}
The spin connection $\omega_{MAB}$ is defined by
\begin{equation}
\omega_{MAB}=\omega^0_{MAB}(e)+T_{MAB},
\end{equation}
where $T_{MAB}$ is the spin torsion tensor.

The transformation laws are
\begin{equation}
\delta e^A_M=-i\kappa{\bar\epsilon}\diamondsuit\Gamma^A\psi_M,
\end{equation}
\begin{equation}
\delta\psi_M=\frac{1}{\kappa}D_M({\hat\omega})\diamondsuit\epsilon+\frac{i}{144}
({\Gamma^{OPQR}}_M -8\Gamma^{PQR}\delta^O_M)\epsilon\diamondsuit{\hat
F}_{OPQR} \equiv \frac{1}{\kappa}{\hat D}_M\epsilon,
\end{equation}
\begin{equation}
\delta A_{MNP}=\frac{3}{2}{\bar\epsilon}\diamondsuit\Gamma_{[MN}\psi_{P]},
\end{equation}
where
\begin{equation}
{\hat F}_{MNPQ}=F_{MNPQ}
-3\kappa{\bar\psi}_{[M}\diamondsuit\Gamma_{NP}\psi_{Q]}.
\end{equation}
We also have
\begin{equation}
{\hat\omega}_{MAB}=\omega_{MAB}+\frac{i\kappa^2}{4}{\bar\psi}_O
\diamondsuit{\Gamma_{MAB}}^{OP}\psi_P.
\end{equation}

In the limit $\vert\tau^{MN}\vert\rightarrow 0$ and
$\beta^M\rightarrow 0$, (\ref{Lagrangian}) reduces to the
CJS eleven-dimensional supergravity Lagrangian~\cite{Julia},
which should be the correct low energy limit of an M-theory. The finiteness
and gauge invariance of the M-theory is guaranteed by the
non-anticommutative field theory~\cite{Moffat3,Moffat4,Moffat5}.
The symmetric tensor $\tau^{MN}$ can be written as
\begin{equation}
\tau^{MN}=\ell^2s^{MN}=\frac{1}{\Lambda^2}s^{MN},
\end{equation}
where $\Lambda$ is a fundamental energy scale chosen to be
$\Lambda=M^{(11)}_{PL}$. Thus, there are no free parameters in
our M-theory.

\section{\bf Supermembranes}

The coordinates of a curved superfield
superspace are given by
\begin{equation}
Z^{\bar M}=(\rho^M,\theta^\alpha)
\end{equation}
where $\theta^\alpha$ denote superfield superspace spinors. We
also introduce the superelfbeins $E_{\bar M}^{\bar A}(Z)$ where
${\bar A}=(A,\alpha)$ are tangent space indices and the pull-back
on the world volume coordinates $\xi^i=(\tau,\sigma_1,\sigma_2)$
(i=0,1,2) is
\begin{equation}
E_i^{\bar A}=\partial_iZ^{\bar M}E_{\bar M}^{\bar A}.
\end{equation}
The supermembrane action using the
$\diamondsuit$-product rule
is given by~\cite{Townsend,Duff}:
\begin{equation}
S_{\rm SM}=\int d^{3}\xi
E\diamondsuit\biggl[-\frac{1}{2}\sqrt{-\gamma}
\gamma^{ij}E_i^{\bar M}\diamondsuit E_j^{\bar
N}\diamondsuit g_{\bar M\bar N}
$$ $$
+\frac{1}{2}\sqrt{-\gamma}
+\frac{1}{3!}\epsilon^{ijk}E_i^{\bar M}\diamondsuit
E_j^{\bar N}\diamondsuit E_k^{\bar P}\diamondsuit A_{\bar M\bar
N\bar P}\biggr]. \end{equation}

The transformation rules are
\begin{equation}
\delta Z^{\bar M}\diamondsuit E^A_{\bar M}=0,\quad \delta
Z^{\bar M}E^\alpha_{\bar
M}={\kappa^\beta(1+\Gamma)^\alpha}_\beta,
\end{equation}
where
$\kappa^\beta(\xi)$ is an anticommuting spacetime spinor and
\begin{equation}
{\Gamma^\alpha}_\beta=\frac{1}{3!}{[\sqrt{-\gamma}\epsilon^{ijk}E_i^{\bar
A}\diamondsuit E_j^{\bar B}\diamondsuit E_k^{\bar C}\Gamma_{\bar
A\bar B\bar C}]^\alpha}_\beta.
\end{equation}
The $\Gamma _a$
are Dirac matrices and $\Gamma_{abc}=\Gamma_{[abc]}$.

The required $\kappa$ symmetry~\cite{Townsend,Duff} is achieved
by certain constraints satisfied by the four-form field strength
$F_{{\bar M}{\bar N}{\bar P}{\bar Q}}$ and the supertorsion.
These constraints amount to saying that the field variables in
the CJS action satisfy the eleven-dimensional supergravity
equations of motion.

\section{\bf M-theory Bosonic Field Equations}

The bosonic action of the M-theory takes the form
\begin{equation}
S_{\rm SGB}=\int
d^{(11)}\rho\sqrt{g^{(11)}}\diamondsuit\biggl[-\frac{1}{2}R_{\diamondsuit}
-\frac{1}{48}F_{MNPQ}\diamondsuit
F^{MNPQ}+\biggl[\frac{\sqrt{2}}{6\cdot(4!)^2}\biggr]\biggl(\frac{1}{\sqrt{g^{(11)}}}\biggr)
$$ $$ \times\diamondsuit
\epsilon^{M_1M_2...M_{11}}F_{M_1M_2M_3M_4}\diamondsuit
F_{M_5M_6M_7M_8}\diamondsuit A_{M_9M_{10}M_{11}}\biggr].
\end{equation}
The metric is $(-+++...+)$, $\epsilon^{0123...}=+1$ and
$F_{MNPQ}=4!\partial_{[M}A_{NPQ]}$ and we have set $8\pi G^{(11)}=c=1$,
where $G^{(11)}$ is the eleven-dimensional gravitational coupling constant.

The field equations are
\begin{equation}
R_{MN}-\frac{1}{2}g_{MN}\diamondsuit R_{\diamondsuit}=-T_{F\,MN},
\end{equation}
\begin{equation}
\label{Fequation}
\frac{1}{\sqrt{g^{(11)}}}\diamondsuit\partial_M(\sqrt{g^{(11)}}\diamondsuit
F^{MNPQ})
=-\biggl[\frac{\sqrt{2}}{2\cdot(4!)^2}\biggr]\biggl(\frac{1}{\sqrt{g^{(11)}}}\biggr)
\diamondsuit\epsilon^{M_1...M_8NPQ}
$$ $$ \times
F_{M_1M_2M_3M_4}\diamondsuit
F_{M_5M_6M_7M_8},
\end{equation}
where
\begin{equation}
\label{Ftensor}
T_{F\,MN}=\frac{1}{48}(8F_{MPQR}\diamondsuit {F_N}^{PQR}-g_{MN}\diamondsuit
F_{SPQR}\diamondsuit F^{SPQR}).
\end{equation}

The bosonic sector of the supermembrane action has
the form
\begin{equation}
\label{bosesmaction}
S_{\rm SMB}=\int d^{3}\xi
\biggl[-\frac{1}{2}\sqrt{-\gamma}\gamma^{ij}\partial_i \rho^M\partial_j
\rho^Ng_{MN}(\rho)
$$ $$
+\frac{1}{2}\sqrt{-\gamma}
+\frac{1}{3!}\epsilon^{ijk}\partial_i\rho^M\partial_j
\rho^N\partial_k\rho^PA_{MNP}(\rho)\biggr].
\end{equation}

\section{Cosmological Solutions}

For cosmological purposes, we shall restrict our attention to an
eleven-dimensional metric of the form
\begin{equation}
g_{MN}=\left(\begin{array}{ccc}
        -1&0&0\\
        0 & a_4^2(t){\tilde g}_{rs}&0\\
        0 & 0 & a_7^2(t){\tilde {g}}_{\Delta\Sigma}
        \end{array}\right).
\end{equation}
Here, $\tilde{g}_{rs}$ (r,s=1,2,3) and
$\tilde{g}_{\Delta\Sigma}$ $(\Delta,\Sigma=5,...,11)$ are the
maximally symmetric three and seven spacelike spaces,
respectively, and $a_4$ and $a_7$ are the corresponding time
dependent cosmological scale factors. We have assumed for
simplicity that the seven extra spacelike dimensions form a
maximally symmetric space, although there is no a priori reason
that this be the case. We shall assume that we are working at
energy scales well below the eleven-dimensional Planck mass,
$M^{(11)}_{\rm PL}$, so that the Grassmann coordinates $\beta^M$
are small compared to $x^M$, $\rho^M\approx x^M$.

The non-vanishing components of the Christoffel symbols are
\begin{equation}
\Gamma^0_{rs}=\frac{{\dot a}_4}{a_4}g_{rs},\quad
\Gamma^0_{\Delta\Sigma}=\frac{{\dot a}_7}{a_7}g_{\Delta\Sigma},\quad
\Gamma^r_{s0}=\frac{{\dot a}_4}{a_4}{\delta^r}_s,
$$ $$
\Gamma^\Gamma_{\Delta 0}=\frac{{\dot
a}_7}{a_7}{\delta^\Gamma}_\Delta,\quad
\Gamma^r_{st}={\tilde\Gamma}^r_{st},\quad
\Gamma^\Gamma_{\Delta\Sigma}={\tilde\Gamma}^\Gamma_{\Delta\Sigma},
\end{equation}
where $g_{rs}=a^2_4{\tilde g}_{rs}$,
$g_{\Delta\Sigma}=a^2_7{\tilde g}_{\Delta\Sigma}$
and
${\tilde\Gamma}^r_{st}$ and
${\tilde\Gamma}^\Gamma_{\Delta\Sigma}$ are the Christoffel
symbols formed from the ${\tilde g}_{rs}$,
${\tilde g}_{\Delta\Sigma}$ and their derivatives.

The non-vanishing components of
the Ricci tensor are
\begin{equation}
\label{Ricci}
R_{00}=3\frac{{\ddot a_4}}{a_4}+7\frac{{\ddot a_7}}{a_7},
$$ $$
R_{rs}=-\biggl[\frac{2k_4}{a_4^2}
+\frac{d}{dt}\biggl(\frac{{\dot a}_4}{a_4}\biggr)+\biggl(3\frac{{\dot
a}_4}{a_4}+7\frac{{\dot a}_7}{a_7}\biggr)\frac{{\dot
a}_4}{a_4}\biggr]g_{rs},
$$ $$
R_{\Delta\Sigma}=-\biggl[\frac{2k_7}{a^2_7}+\frac{d}{dt}\biggl(\frac{{\dot
a}_7}{a_7}\biggr)+\biggl(3\frac{{\dot a}_4}{a_4}+7\frac{{\dot
a}_7}{a_7}\biggr)\frac{{\dot a}_7}{a_7}\biggr]g_{\Delta\Sigma},
\end{equation}
where $k_4$ and $k_7$ are the curvature constants of
four-dimensional and seven-dimensional space. Positive and negative values
of $k_4$ and $k_7$ correspond to the sphere and the pseudosphere,
respectively, while vanishing values of $k_4$ and $k_7$ correspond to flat
spaces. The energy-momentum tensor can be expressed in a perfect
fluid form
\begin{equation}
T_{MN}={\rm diag}(-\rho, pg_{rs}, p'g_{\Delta\Sigma}),
\end{equation}
where $\rho=\rho_m+\rho_F, p=p_m+p_F$ and $p'=p'_m+p'_F$.

We now adopt the Freund-Rubin ansatz for which all components of the
four-form field $F_{MNPQ}$ vanish except~\cite{Freund,Freund2}:
\begin{equation}
F_{\mu\nu\rho\sigma}=mf(t)\frac{1}{\sqrt{-g^{(4)}}}\epsilon_{\mu\nu\rho\sigma},
\end{equation}
where $\mu,\nu=0,1,2,3$ and $m$ is a constant.
With this ansatz, the trilinear contributions in $A_{MNP}$ and
its derivatives in the action vanish. The three form potential
field $A_{MNP}$ is required to live on a three manifold $M_3$
(i.e. to be a maximally form-invariant tensor on $M_3$) :
\begin{equation}
A_{MNP}\equiv A_{rst}=mA(t)\sqrt{g^{(3)}}\epsilon_{rst},
\end{equation}
where $A(t)$ is a function of time and $g^{(3)}$ and
$\epsilon_{rst}$ are the metric determinant and the Levi-Civita
symbol on $M_3$, respectively, and we have $f(t)=-{\dot A}(t)$.

Let us assume that the four-form tensor
$F_{MNPQ}$ dominates, so that we neglect the effects of the
matter fields with density $\rho_m$. We shall use
(\ref{diamondproduct}) to expand the products of the F-tensors
in small values of $\vert\tau^{MN}\vert$. We neglect the
contributions from the $\diamondsuit$-products of the metric
$g_{MN}$ and its derivatives compared to the
$\diamondsuit$-products of the F-tensors. Using the results that
$\epsilon_{\mu\alpha\rho\sigma}{\epsilon_\nu}^{\alpha\rho\sigma}
=6m^2f^2(t)g_{\mu\nu}(-g^{(4)})$ and
$\epsilon_{\mu\nu\rho\sigma}\epsilon^{\mu\nu\rho\sigma}=24m^2f^2(t)(-g^{(4)})$,
we find to first order in $\vert\tau\vert$:
\begin{equation}
T_{F\,MN}=\frac{1}{2}\epsilon
m^2\biggl(f^2-\frac{{\dot f}^2}{2\Lambda^2}\biggr)g_{MN},
\end{equation}
where we have chosen
\begin{equation}
\tau^{00}=-\frac{1}{\Lambda^2}.
\end{equation}
Moreover, $\epsilon =+1$ for $M,N=\mu,\nu$ and $-1$ for
$M,N=\Delta,\Sigma$. Eq.(\ref{Fequation}) becomes
\begin{equation}
\frac{d}{dt}\biggl[(a_7(t))^7f(t)\biggr]=0.
\end{equation}

Our spontaneous compactification leads to
\begin{equation}
\label{R1} R_{\mu\nu}-\frac{1}{2}g_{\mu\nu}R=\lambda_4g_{\mu\nu},
\end{equation} and
\begin{equation}
\label{R2}
R_{\Delta\Sigma}-\frac{1}{2}g_{\Delta\Sigma}R=\lambda_7g_{\Delta\Sigma},
\end{equation}
where
\begin{equation}
\lambda_4=-\frac{1}{2}m^2\biggl(f^2-\frac{{\dot
f}^2}{2\Lambda^2}\biggr),\quad \lambda_7
=\frac{1}{2}m^2\biggl(f^2-\frac{{\dot f}^2}{2\Lambda^2}\biggr).
\end{equation}

Let us assume that
\begin{equation}
\label{Cconstant}
C\approx f^2-\frac{{\dot f}^2}{2\Lambda^2},
\end{equation}
where $C$ is a constant. In the
limit, $\Lambda\rightarrow\infty$, we obtain the standard supersymmetric
vacuum result
\begin{equation}
\lambda_4=-\frac{1}{2}m^2f^2,\quad \lambda_7=\frac{1}{2}m^2f^2.
\end{equation}
The eleven-dimensional space becomes a product of a four-dimensional
Einstein AdS space with negative
cosmological constant and a seven-dimensional Einstein space with positive
cosmological constant~\cite{Duff2}.
If we require the vacuum to be supersymmetric by
demanding covariantly constant spinors $\theta$ for which
\begin{equation}
\delta\psi={\bar D}_M\theta=0,
\end{equation}
where
\begin{equation}
{\bar D}_M=D_M+\frac{i\sqrt{2}}{288}(\Gamma^{NPQR}_M-8\Gamma^{PQR}
{\delta_M}^N)F_{NPQR},
\end{equation}
then for $m\not= 0$, $N=8$ supersymmetry uniquely chooses $AdS\times S^7$
with an $SO(8)$-invariant metric on $S^7$~\cite{Englert,Duff2}.

If we choose
$f^2={\dot f}^2/2\Lambda^2$, we get $\lambda_4=\lambda_7=0$ and flat
$(3+1)$ and seven-dimensional spaces. On the other hand, if we choose $f^2
< {\dot f^2}/2\Lambda^2$, then $\lambda_4 > 0$ and $\lambda_7 < 0$ and we
obtain a positive cosmological constant in $(3+1)$ spacetime,
corresponding to a dS universe, and a seven-dimensional space
with negative curvature.

From (\ref{R1}) and (\ref{R2}), we obtain
\begin{equation}
\label{a1}
3{\ddot a}_4+7\frac{{\ddot a}_7a^2_4}{a_7}=\lambda_4a^2_4,
\end{equation}
\begin{equation}
\label{a2}
2k_4+{\ddot a}_4a_4+2{\dot a}^2_4+7\frac{{\dot a}_7{\dot
a}_4a_4}{a_7}=\lambda_4a^2_4,
\end{equation}
\begin{equation}
2k_7+{\ddot a}_7a_7+6{\dot a}^2_7
+3\frac{{\dot a}_4{\dot a}_7a_7}{a_4}=\lambda_7a^2_7.
\end{equation}
Combining (\ref{a1}) and (\ref{a2}) gives
\begin{equation}
\biggl(\frac{{\dot a}_4}{a_4}\biggr)^2+\frac{k_4}{a^2_4}
-\frac{7}{6}\frac{{\ddot a}_7}{a_7}
+\frac{7}{2}\frac{{\dot a}_7{\dot a}_4}{a_4a_7}=\frac{1}{3}\lambda_4.
\end{equation}
For solutions in which the scale factor $a_4$ expands
faster than $a_7$, we get the standard four-dimensional Friedmann equation:
\begin{equation}
\biggl(\frac{{\dot a}_4}{a_4}\biggr)^2
+\frac{k_4}{a^2_4}=\frac{1}{3}\lambda_4.
\end{equation}
This has the dS inflationary solution for $k_4=0$ and $\lambda_4
> 0$:
\begin{equation}
a_4=B\exp\biggl(\sqrt{\frac{\lambda_4}{3}}t\biggr),
\end{equation}
where $B$ is a constant.

Let us now consider a time-dependent solution
including matter density $\rho_m$, obtained from additional matter fields generated
by our compactification to a four-dimensional spacetime. We write
\begin{equation}
\label{4-Einstein}
R_{\mu\nu}-\frac{1}{2}g_{\mu\nu}R=-T_{\mu\nu},
\end{equation}
where
\begin{equation}
T_{\mu\nu}=T_{F\mu\nu}+T_{M\mu\nu},
\end{equation}
and
\begin{equation}
T_{F\mu\nu}=\frac{1}{2}m^2S(t)g_{\mu\nu}.
\end{equation}
$T_{M\mu\nu}$ denotes the energy-momentum
contribution in four dimensions from the matter fields.
We also have
\begin{equation}
S=f^2-\frac{{\dot f}^2}{2\Lambda^2}.
\end{equation}

The $T_{M\mu\nu}$ can be written for a comoving cosmological
fluid in the form
\begin{equation}
T_{M\mu\nu}={\rm diag}(-\rho_m, p_m, p_m, p_m),
\end{equation}
where the $\rho_m$ and $p_m$ denote the density and pressure of matter fields,
respectively. We rewrite (\ref{4-Einstein}) as
\begin{equation}
R_{\mu\nu}=-{\tilde T}_{\mu\nu},
\end{equation}
\begin{equation}
{\tilde
T}_{\mu\nu}=T_{\mu\nu}-\frac{1}{2}g_{\mu\nu}{T_\lambda}^\lambda.
\end{equation} We express the energy-momentum tensor ${\tilde
T}_{\mu\nu}$ as an effective perfect fluid
\begin{equation}
{\tilde T}_{\mu\nu}=(\rho+p)u_\mu
u_\nu+\frac{1}{2}(\rho-p)g_{\mu\nu},
\end{equation}
where the
four-velocity has the components $u^0=1$, $u^m=0$ and
$\rho=\rho_m+\rho_F$. We have
\begin{equation}
\rho_F=-\frac{3}{2}m^2S,\quad p_F=\frac{3}{2}m^2S,
\end{equation}
so that $\rho_F=-p_F$ which is the equation of state for the vacuum density
$\rho_{\rm vac}$.

From (\ref{Ricci}) we obtain
\begin{equation}
\label{modfriedmann}
\biggl(\frac{{\dot a}_4}{a_4}\biggr)^2+\frac{k_4}{a_4^2}
=\frac{1}{3}\rho+\frac{1}{3}\lambda_4-\frac{7}{2}\frac{{\dot
a}_7}{a_7}\frac{{\dot a}_4}{a_4} +\frac{7}{6}\frac{{\ddot
a}_7}{a_7}, \end{equation}
\begin{equation}
\label{acceleration}
\frac{{\ddot a}_4}{a_4}
=-\frac{1}{6}(\rho+3p)+\frac{1}{3}\lambda_4-\frac{7}{3}\frac{{\ddot
a_7}}{a_7}, \end{equation}
\begin{equation}
\frac{{\ddot a}_7}
{a_7}+\frac{2k_7}{a_7^2}
=-\biggl[6\biggl(\frac{{\dot a}_7^2}{a_7^2}\biggr)
+3\frac{{\dot a}_4{\dot
a}_7}{a_4a_7}\biggr].
\end{equation}

Let us again assume that $f$ and ${\dot f}$ are slowly varying
and choose the solution $\lambda_4>0$, $f^2<{\dot f}^2$
and $\rho_F=-p_F$. We also assume the equation of
state $p_m=w\rho_m$ and that for the present universe $w\approx 0$.
Then, we have for a four-dimensional universe expanding faster
than the seven-dimensional compact space:
\begin{equation}
\label{newfriedmann}
\biggl(\frac{{\dot a}_4}{a_4}\biggr)^2+\frac{k_4}{a_4^2}=
\frac{1}{3}\rho_m + \frac{1}{3}\lambda_4
\end{equation}
\begin{equation}
\label{accel}
\frac{{\ddot a}_4}{a_4}
=-\frac{1}{6}(\rho_m+3p_m)+\frac{1}{3}\lambda_4.
\end{equation}
We can write (\ref{newfriedmann}) in the form
\begin{equation}
\Omega_M+\Omega_{k_4}+\Omega_{\lambda_4}=1,
\end{equation}
where
\begin{equation}
\Omega_M=\frac{\rho_m}{3H^2},\quad
\Omega_{k_4}=-\frac{k_4}{a_4^2H^2},\quad
\Omega_{\lambda_4}=\frac{\lambda_4}{3H^2}.
\end{equation}
By choosing
\begin{equation}
\Omega_M=0.32,\quad \Omega_{k_4}=0,\quad \Omega_{\lambda_4}=0.68,
\end{equation}
we can obtain a fit to all the current observational
data~\cite{Perlmutter,Netterfield}.

Generally, phenomenological matter contributions are not
permitted in the eleven-dimensional M-theory supergravity,
because of the restrictions incurred by the generalized
supersymmetric invariance of the action. However, when we perform our
spontaneous compactification the fermionic gravitino
contributions and the elvbeins will generate matter fields in the $M_4\times M_7$
product space, which can describe the `{\it dark matter}' and the baryons and
neutrinos. The F-field contributions, on the other hand, describe the `{\it dark
energy}' associated with the vacuum and the cosmological constant.

\section{\bf A Realistic M-Theory of Particles?}

We now turn to the important question as to whether we can derive a
realistic description of particle unification in our M-theory.
Early attempts to obtain an isometry group from the $M^{pqr}$
spaces introduced by Witten~\cite{Witten4} failed to give a
realistic descriptions of quarks and leptons~\cite{Duff2}.
Witten starts from the eleven-dimensional model and assumes that
the seven-dimensional internal space is a coset space $G/H$.
Choosing the minimal $H$ with $G=SU(3)\times SU(2)\times U(1)$
allows a nontrivial action of each of the factors $SU(3), SU(2)$
and $U(1)$ of $G$ on $G/H$ and this space is exactly
seven-dimensional. Decomposing $g_{MN}$ into the product
space $M_4\times M_7$, one finds that the action can gauge
$SU(3)\times SU(2)\times U(1)$ in $D=4$. However, the
lepton-quark part of the fermion decomposition does not fit into
the required $SU(3)\times SU(2)\times U(1)$ group
representations.

Another attempt to obtain a description of particles consists of a spontaneous
compactification of the type using the Freund-Rubin ansatz for
the four-form field $F_{MNPQ}$ and a round seven sphere $S^7$ or
squashed seven sphere~\cite{Englert}. Such a compactification
does not permit singularities in the smooth internal space $S^7$ and therefore cannot
yield a chiral fermion spectrum and non-Abelian gauge fields within $SO(8)$, which
does not contain the standard model group $SU(3)\times SU(2)\times U(1)$.

Some time ago it was
speculated that some of the gauge bosons and fermions are composite as in the model
of Ellis, et al.~\cite{Ellis}. The expectation that there are two kinds of gauge
bosons, the ``elementary'' and the ``composite'' is a consequence of Kaluza-Klein
theories and not due to an ad hoc assumption. Kaluza-Klein theories unify gravity
and Yang-Mills theories in two ways. First D-dimensional general covariance of our
superspace $\rho^M\rightarrow \rho^{'M}(\rho)$ and then secondly D-dimensional local
super Lorentz invariance $\delta e_M^A(\rho)=L^A_B(\rho)e^B_M(\rho),
\delta\omega^{AB}_M=-D_ML^{AB}(\rho)$ give in eleven dimensions a superspace
orthogonal group. Thus, the elfbeins $e^A_M$ are the gauge bosons of superspace
general covariance and the superspace spin connections are the gauge bosons of
super Lorentz invariance. Whereas the $e^A_M$ are the ``elementary'' fields, the
$\omega^{AB}_M$ are the ``composite'' fields. Since the $\omega^{AB}_M$ have no
kinetic terms of their own, we have to express them in terms of the elfbein and its
derivatives. The Kaluza-Klein compactification gives rise to both elementary and
composite gauge bosons in $D=4$.

At energies well below the Planck mass, our superspace can be
approximated by the ordinary four-dimensional spacetime
coordinates $\rho^\mu\sim x^\mu$, and the elementary gauge
bosons $A_\mu(x)$ come from the the $e^a_\mu (x,y)$ and
correspond to a gauge group given by the isometry group of the
extra dimensions, $SO(8)$ for the round $S^7$, while the
composite gauge fields $C_\mu(x)$ come from the
$\omega^{ab}_\mu(x,y)$ and correspond to the tangent space group
of the extra internal dimensions, $SO(7)$, for the case of $d=7$.
Cremmer and Julia~\cite{Cremmer} showed that when the three form
field $A_{MNP}$ is accounted for, the hidden symmetry  could be
larger than $SO(7)$ and may be as large as $SO(8)$ or $SU(8)$.
Now there are many ways to embed $SU(3)\times SU(2)\times U(1)$
in `elementary' $SO(8)\times$ `composite' $SU(8)$. However, there
still remain unresolved questions about the physical properties
of these bound state representations of gauge bosons and
fermions, and whether they permit chiral representations in four dimensions.

Perhaps the most promising way to obtain a realistic description of particles in
four dimensions from a compactification of M-theory, with complex chiral
fermion representations and non-abelian gauge fields, is to seek
a compactification on a seven manifold that contains singularities. As shown by
Acharya and Witten~\cite{Acharya}, we should seek a hyper-Kahler manifold
compactification with conical singularies that generate the desired chiral fermions
and non-Abelian gauge fields within a four-dimensional supersymmetric unified model,
which contains the standard $SU(3)\times SU(2)\times U(1)$ group.  Such
a compactification scheme could be generalized to our supermanifold noncommutative
and non-anticommutative geometry.

\section{Conclusions}

Our M-theory describes a finite quantum field theory in an
eleven-dimensional super manifold in which the action is constructed from a
$\diamondsuit$-product of field operators based on eleven-dimensional CJS
supergravity theory. The quantum gravity and quantum gauge field parts of
the action will be finite for $\Lambda < \infty$ due to the finiteness of
the non-anticommutative quantum field theory. In the limits
$\Lambda\rightarrow\infty, \beta^M\rightarrow 0$ and $\vert\tau^{MN}\vert\rightarrow
0$, we obtain the low energy limit of eleven-dimensional CJS supergravity. The
compactified version of this theory has the same massless ten-dimensional particle
spectrum as type-IIA superstring theory and is connected to the latter theory by a
duality transformation.

The M-theory eleven-dimensional field equations are invariant under
generalized $\diamondsuit$-product supersymmetric gauge transformations,
which can be thought of as classical deformations of the standard
supersymmetric gauge transformations of CJS supergravity. The generalized
$\diamondsuit$-product supersymmetric vacuum leads to de Sitter
space solutions and thus provides an empirical basis for a
realistic cosmology. The no-go theorems
in~\cite{Gibbons,Maldacena} do not apply in this case, because
they are derived from standard supersymmetric theories. Our
M-theory with generalized supersymmetric equations predicts an
inflationary period in the early universe. As the universe
expands, $f^2$ can tend towards ${\dot f}^2/2\Lambda^2$ and
produce a small four-dimensional cosmological constant with
$\lambda_4 > 0$ and an accelerating universe at present. We have
described the dominant dark matter and dark energy by the
energy-momentum matter field tensor and the
four-form F-field energy-momentum tensor, respectively.

The problem of chiral fermions can be resolved by permitting isolated
singularities in our seven-dimensional compact space $M_7$. To
obtain a realistic mass spectrum, which contains the standard
model $SU(3)\times SU(2)\times U(1)$, we should seek a compactification on a
seven-dimensional hyper-Kahler type manifold and generate four-dimensional dS
solutions through our generalized supersymmetric cosmological solutions.

\vskip 0.2 true in
{\bf Acknowledgments}
\vskip 0.2 true in
I thank Michael Duff, Bobby Acharya, Lee Smolin, Michael Clayton and Pierre Savaria
for helpful and stimulating conversations. This work was supported by the Natural
Sciences and Engineering Research Council of Canada.
\vskip 0.5 true in

\end{document}